\documentclass[10pt,prc,twocolumn,
nofootinbib,
noshowkeys,
noshowpacs,
superscriptaddress,
floatfix
]{revtex4-2}
\usepackage{amsmath,amsfonts,amsthm,amssymb}
\usepackage[dvips]{graphics,graphicx}
\usepackage[usenames,dvipsnames]{color}
\definecolor{darkblue}{RGB}{0,0,196}
\definecolor{darkgreen}{RGB}{0,120,0}
\usepackage[colorlinks=true,linktocpage=true,linkcolor=darkblue,citecolor=red,urlcolor=darkblue]{hyperref}
\usepackage{cancel}
\usepackage{bbold}
\usepackage{multirow}
\usepackage{longtable}
\usepackage{color}
\usepackage[normalem]{ulem}
\usepackage{hyperref}
\usepackage{bigints}
\usepackage{xparse}
\usepackage{physics}
\usepackage{verbatim}
\usepackage{minibox}
\usepackage{comment}
\usepackage{appendix}
\usepackage{slashed}
\usepackage{marginnote}
\usepackage{graphicx}
\usepackage[nice]{nicefrac}
\usepackage{amsmath}
\usepackage{hepunits}
\include{commands}
\usepackage{stackengine,scalerel}
\newcommand\hstar[1]{\ThisStyle{\ensurestackMath{%
  \setbox0=\hbox{$\SavedStyle#1$}%
  \stackengine{0pt}{\copy0}{\kern.2\ht0\smash{\SavedStyle\star}}{O}{c}{F}{T}{S}}}}

\definecolor {darkgreen}{rgb}{0.2,0.7,0.2}

\begin{document}
\title{Canonical and phenomenological formulations of spin hydrodynamics}

\author{Asaad Daher}
\email{asaad.daher@ifj.edu.pl}
\affiliation{Institute  of  Nuclear  Physics  Polish  Academy  of  Sciences,  PL-31-342  Krak\'ow,  Poland}
\author{Arpan Das}
\email{arpan.das@ifj.edu.pl}
\affiliation{Institute  of  Nuclear  Physics  Polish  Academy  of  Sciences,  PL-31-342  Krak\'ow,  Poland}

\author{Wojciech Florkowski}
\email{wojciech.florkowski@uj.edu.pl}
\affiliation{Institute of Theoretical Physics, Jagiellonian University, PL-30-348 Krak\'ow, Poland}

\author{Radoslaw Ryblewski}
\email{radoslaw.ryblewski@ifj.edu.pl}
\affiliation{Institute  of  Nuclear  Physics  Polish  Academy  of  Sciences,  PL-31-342  Krak\'ow,  Poland}

\begin{abstract}
Two formulations of relativistic hydrodynamics of particles with spin 1/2 are compared. The first approach, dubbed the canonical one, uses expressions for the energy-momentum and spin tensors that have properties that follow a direct application of Noether's theorem, which yields a totally antisymmetric spin tensor. The other one is based on a simplified form of the spin tensor and is commonly used in the current literature under the name of a phenomenological approach. We show that these two frameworks are equivalent, i.e., they can be directly connected by a suitably defined pseudogauge transformation, only if the first framework is initially improved by a suitable modification of the energy-momentum tensor (addition of a divergence-free term that cannot be interpreted as a pseudogauge). Our analysis uses arguments related to the positivity of entropy production. The latter turns out to be equivalent for the improved canonical and phenomenological frameworks.
\end{abstract}

\date{\today \hspace{0.2truecm}}

\maketitle
\flushbottom






%
\section{Introduction}
%
Recent observations of spin polarization of weakly decaying hyperons produced in relativistic heavy-ion experiments across various collision energies~\cite{STAR:2017ckg, STAR:2018gyt,STAR:2019erd,ALICE:2019onw,ALICE:2019aid,STAR:2020xbm,Kornas:2020qzi,STAR:2021beb,ALICE:2021pzu} have provided a unique probe to study polarization phenomena in relativistic nuclear matter under rotation \cite{lisa2021}. Motivated by the earlier successes of relativistic fluid dynamics in heavy-ion phenomenology \cite{Florkowski:2017olj}, several extensions of relativistic hydrodynamics for the spin-polarized fluids have been developed using quantum statistical density operators~\cite{Becattini:2007nd,Becattini:2009wh,Becattini:2012pp,Becattini:2018duy,Hu:2021lnx}, relativistic kinetic theory~\cite{Florkowski:2017ruc,Florkowski:2017dyn,Florkowski:2018ahw,Florkowski:2018fap,Florkowski:2019qdp,Bhadury:2020puc,Bhadury:2020cop,Speranza:2020ilk,Weickgenannt:2020aaf,Weickgenannt:2021cuo,Shi:2020htn,Peng:2021ago,Sheng:2021kfc}, effective Lagrangian approach~\cite{Montenegro:2017rbu,Montenegro:2017lvf,Montenegro:2018bcf,Montenegro:2020paq}, entropy current analysis~\cite{Hattori:2019lfp,Fukushima:2020ucl,Li:2020eon,She:2021lhe,Hongo:2021ona,Daher:2022wzf,Biswas:2022bht,Biswas:2023qsw,Wang:2021ngp}, holography~\cite{Gallegos:2020otk,Garbiso:2020puw} and equilibrium partition functions~\cite{Gallegos:2021bzp}.

At the macroscopic level, relativistic hydrodynamics with spin can be formulated on the grounds of the conservation laws of conserved currents, i.e., energy-momentum tensor, baryon current, and the total angular momentum current, with the last one composed of spin and orbital parts. In special relativity, however, the decomposition of the total angular momentum current obtained by choosing a particular form of the energy-momentum tensor and the spin tensor is not unique. A given pair of these currents can always be transformed to another one through the so-called pseudo-gauge \cite{HEHL197655,Speranza:2020ilk}. Of particular importance in this context is the \emph{Belinfante pseudo-gauge} based on Belinfante’s improved, i.e. symmetric, energy-momentum tensor \cite{BELINFANTE1939887,BELINFANTE1940449,Rosenfeld1940}, which provides a natural connection to Einstein’s relativity theory. Another choice, known as the \emph{phenomenological pseudo-gauge}, deals with an arbitrary energy-momentum tensor, i.e., with symmetric and antisymmetric components, and the spin tensor that is antisymmetric in the last two indices, which turns out to be especially convenient for the construction of hydrodynamic frameworks. Last but not least, at the microscopic level, the \emph{canonical} forms of conserved currents with the energy-momentum tensor having both symmetric and antisymmetric parts and the spin tensor being totally antisymmetric in all its Lorentz indices, can be obtained. Such a choice of conserved currents is motivated from the standpoint of the quantum field theory of Dirac fermions when applying the Noether theorem \cite{Peng:2021ago}.

While the change to a particular pseudo-gauge leaves the total charges and dynamical equations intact, it modifies the densities of the macroscopic quantities, hence introducing possible pseudo-gauge dependence of the resulting hydrodynamic formalism~\cite{Florkowski:2018fap}. Moreover, in local thermal equilibrium, the choice of decomposition of the angular momentum into an orbital and a spin part can significantly affect the predictions of spin polarization \cite{Buzzegoli:2021wlg}.  To explore such freedom, in Ref.~\cite{Fukushima:2020ucl} the dissipative spin hydrodynamics was formulated using Belinfante pseudo-gauge, starting from the phenomenological set of energy-momentum and spin tensors.

In the present work, we formulate the dissipative spin hydrodynamics using the entropy current analysis and canonical forms of conserved currents. We find that the totally antisymmetric nature of the spin tensor introduces the new properties into the hydrodynamic description that are otherwise missing in the phenomenological~\cite{Hattori:2019lfp} or the Belinfante~\cite{Fukushima:2020ucl} frameworks. In particular, our work shows that the naïve use of arbitrary forms of canonical currents conflicts with the principle of entropy production. To resolve this problem, we construct an \emph{improved} form of the canonical energy-momentum tensor without affecting the conservation equations. We prove that the resulting improved canonical definitions can be considered a suitable choice for a first-order dissipative hydrodynamic framework. Using proper pseudo-gauge transformation, we show an equivalence between the improved canonical framework and the phenomenological framework of dissipative spin hydrodynamics.
\smallskip

The paper is organized as follows. We begin our discussion of the canonical formulation of first-order dissipative spin hydrodynamics in Sec.~\ref{sec:2}. In this section, we argue that a naïve use of a general canonical currents is in conflict with the principle of entropy production for a dissipative system. In Sec.~\ref{sec:3} we discuss an improved canonical framework in which one can uniquely identify different dissipative currents using entropy current analysis. In Sec.~\ref{sec:4} we show that one can get the phenomenological framework of dissipative spin hydrodynamics starting from the improved canonical framework, while in Sec.~\ref{sec:5} we justify the derivation of the improved canonical framework from the phenomenological one. In Sec.~\ref{sec:6} we summarize and conclude.   

\section{Entropy current analysis: canonical framework}
\label{sec:2}
%
For a spin-polarized fluid, the hydrodynamic framework can be constructed based on the conservation of energy-momentum tensor $T^{\mu\nu}$ and total angular momentum tensor $J^{\mu\alpha\beta}$\footnote{Here, for simplicity, we assume there are no other conserved charges in the system.} 
\begin{align}
    & \partial_{\mu}T^{\mu\nu}=0, \label{equ1pap1}\\
    & \partial_{\mu}J^{\mu\alpha\beta} =\partial_{\mu}S^{\mu\alpha\beta} +2 T^{[\alpha\beta]}=0.
     \label{equ2pap1}
\end{align}
For symmetric and antisymmetric part of arbitrary tensor $X^{\mu\nu}$ we use the notation $X^{\mu\nu}_{(s)}\equiv X^{(\mu\nu)}=(X^{\mu\nu}+X^{\nu\mu})/2$ and $X^{\mu\nu}_{(a)}\equiv X^{[\mu\nu]}=(X^{\mu\nu}-X^{\nu\mu})/2$, respectively. In Eq.~\eqref{equ2pap1} we have taken advantage of the fact that the total angular momentum $J^{\mu\alpha\beta}$ can be decomposed into an orbital angular momentum component $L^{\mu\alpha\beta}$, which can be expressed using the energy-momentum tensor as $L^{\mu\alpha\beta}=2 x^{[\alpha}T^{\mu\beta]}$, and the intrinsic angular momentum (spin) component $S^{\mu\alpha\beta}$.

The orbital and the spin parts, with the latter being given by the fully antisymmetric tensor, can be argued to originate from using the Noether theorem for the Lorentz symmetry of a non-scalar field theory, namely, for our interest, the Dirac field. Herein, we demand the Lorentz structure of the spin tensor to be preserved at the macroscopic level. To this end, we define the constitutive relations for the energy-momentum tensor and the totally antisymmetric spin tensor up to the first order in the gradient expansion as
\begin{align}
    & T^{\mu\nu}_{\text{can}} = T^{\mu\nu}_{(0)}+T^{\mu\nu}_{\text{can}(1)}, \label{equ3pap1}\\
    & S^{\mu\alpha\beta}_{\text{can}}=u^{\mu}S^{\alpha\beta}+u^{\beta}S^{\mu\alpha}+u^{\alpha}S^{\beta\mu} +S^{\mu\alpha\beta}_{\text{can}(1)}. \label{equ4pap1}
\end{align} 
The leading-order contribution to $T^{\mu\nu}_{\text{can}}$ in Eq.~\eqref{equ3pap1} has the perfect-fluid form
\begin{align} &T^{\mu\nu}_{(0)}=\varepsilon u^{\mu}u^{\nu}-p\Delta^{\mu\nu},
\end{align}
where $\varepsilon$ and $p$ are the energy density and pressure, respectively, $u^{\mu}$ is the fluid four velocity satisfying the normalization condition $u^{\mu}u_{\mu}=1$, $\Delta^{\mu\nu}\equiv g^{\mu\nu}-u^{\mu}u^{\nu}$ is the symmetric operator projecting onto the space orthogonal to $u^{\mu}$, i.e. $\Delta^{\mu\nu}u_{\mu}=0$, and $g_{\mu\nu}=\text{diag}(+1,-1,-1,-1)$ is the Minkowski metric. The spin density tensor $S^{\mu\nu}$ is antisymmetric, i.e. $S^{\mu\nu}=-S^{\nu\mu}$, therefore the leading-order part of the spin tensor $S^{\mu\alpha\beta}_{\text{can}}$ is totally antisymmetric. The terms $T^{\mu\nu}_{\text{can}(1)}$ and $S^{\mu\alpha\beta}_{\text{can}(1)}$ denote the first-order derivative corrections to the energy-momentum tensor and the spin tensor, respectively. In general, $T^{\mu\nu}_{\text{can}(1)}$ contains both a symmetric $T^{\mu\nu}_{\text{can}(1s)}$ and an antisymmetric $T^{\mu\nu}_{\text{can}(1a)}$ part. We can assume that, similarly to the leading-order part, the $S^{\mu\alpha\beta}_{\text{can}(1)}$ part is totally antisymmetric, although this does not affect our analysis.
The tensor $S^{\alpha\beta}$ plays a role analogous to the number density in the presence of a conserved charge~\cite{Hattori:2019lfp,Fukushima:2020ucl}. Near local thermal equilibrium, it satisfies the generalized laws of thermodynamics, namely
\begin{align} \varepsilon+p&=Ts+\omega_{\alpha\beta}S^{\alpha\beta},\nonumber\\ 
    d\varepsilon&=Tds+\omega_{\alpha\beta}dS^{\alpha\beta},    \label{eqthermo} \\ dp&=sdT+S^{\alpha\beta}d\omega_{\alpha\beta},
\nonumber
\end{align}
where $s$ is the entropy density, $T$ is the temperature, and the antisymmetric tensor $\omega^{\mu\nu}$ can be considered as the spin chemical potential conjugate to the spin density $S^{\mu\nu}$~\cite{Hattori:2019lfp}. 

At this point we should comment on the gradient order of the quantities associated with spin. Throughout the manuscript we consider $\omega^{\mu\nu}\sim\mathcal{O}(\partial^1)$, as argued in Refs.~\cite{Hattori:2019lfp,Fukushima:2020ucl}, taking into account that the chemical spin potential at global equilibrium can be expressed by the thermal vorticity tensor $\varpi_{\mu\nu}=- \partial_{[\mu}(u_{\nu]}/T)$. At the same time, $S^{\mu\nu}$ is considered to be $\mathcal{O}(\partial^0)$, which is consistent with the derivative counting of $\omega^{\mu\nu}$ in the high-temperature limit, where one can assume that $S^{\mu\nu}\sim T^2\omega^{\mu\nu}$~\cite{Wang:2021ngp}\footnote{This is not necessarily the only choice. For example, in Ref.~\cite{She:2021lhe}, dissipative spin hydrodynamics was developed, where both the spin chemical potential and the spin density tensor are assumed to be $\mathcal{O}(\partial^0)$.}.

In the presence of the spin chemical potential and spin density, the nonequilibrium entropy current can be generalized to first-order terms in the gradient expansion as follows~\cite{Hattori:2019lfp,Fukushima:2020ucl}\footnote{Note that in addition to the antisymmetry property of $S^{\mu\nu}$ one can also impose the condition that $S^{\mu\nu}$ is orthogonal to $u^{\mu}$, i.e. $S^{\mu\nu}u_{\mu}=0$ ~\cite{Li:2020eon}. This is analogous to the Frenkel condition. This additional condition reduces the number of degrees of freedom of the spin density tensor. In the present study, we do not impose such conditions unless explicitly stated.}.
\begin{align} \mathcal{S}^{\mu}_{\text{can}}& =T^{\mu\nu}_{\text{can}}\,\beta_{\nu}+p\,\beta^{\mu}-\omega_{\alpha\beta}S^{\alpha\beta}\beta^{\mu} +\mathcal{O}(\partial^2)\nonumber\\ & = \mathcal{S}^{\mu}_{(0)}+ T^{\mu\nu}_{\text{can}(1)}\,\beta_{\nu} +\mathcal{O}(\partial^2). \label{equ6pap1}
\end{align}
Here $\mathcal{S}^{\mu}_{(0)}\!=\!s u^{\mu}$ is the equilibrium part of the entropy current, $\beta\equiv 1/T$ is the inverse temperature, and $\beta^{\mu}\equiv \beta u^{\mu}$. Note that up to the first order in the gradient expansion, in the expression of the nonequilibrium entropy current $S^{\mu\alpha\beta}_{\text{can}(1)}$ plays no role. Only the dissipative part of the energy-momentum tensor, i.e. $T^{\mu\nu}_{\text{can}(1)}$ contributes to the nonequilibrium entropy current.

So far, the explicit forms of the various dissipative currents in $T^{\mu\nu}_{\text{can}(1)}$ are not uniquely defined. They can be uniquely determined by imposing the second law of thermodynamics, i.e. $\partial_{\mu}\mathcal{S}^{\mu}_{\text{can}}\geq 0$. Using Eq.~\eqref{equ6pap1}, the divergence of the entropy current can be expressed as
\begin{align} \partial_{\mu}\mathcal{S}^{\mu}_{\text{can}}& =\beta\left[T\partial_{\mu}\mathcal{S}^{\mu}_{(0)}+u_{\nu}\partial_{\mu}T^{\mu\nu}_{\text{can}(1)}\right] +T^{\mu\nu}_{\text{can}(1)}\partial_{\mu}\beta_{\nu}+\mathcal{O}(\partial^3).
\label{equ7pap1}
\end{align}

Using the conservation equations of the canonical energy-momentum tensor and the total angular momentum tensor, Eq.~\eqref{equ7pap1} can be simplified. 
The conservation of the energy-momentum tensor~\eqref{equ1pap1} implies 
\begin{align} & u_{\nu}\partial_{\mu}T^{\mu\nu}_{\text{can}(1)} = -T\partial_{\mu}\mathcal{S}^{\mu}_{(0)}-\omega_{\alpha\beta}\partial_{\mu}(u^{\mu}S^{\alpha\beta}), \label{equ8pap1}
\end{align}
which allows us to write Eq.~\eqref{equ7pap1} in the following way
\begin{align} \partial_{\mu}\mathcal{S}^{\mu}_{\text{can}}& = -\beta\omega_{\alpha\beta}\partial_{\mu}(u^{\mu}S^{\alpha\beta})+T^{\mu\nu}_{\text{can}(1)}\partial_{\mu}\beta_{\nu}+\mathcal{O}(\partial^3). \label{equ9pap1}
\end{align}
Furthermore, using the constitutive relation for the canonical spin tensor given in Eq.~\eqref{equ4pap1}, the conservation of total angular momentum~\eqref{equ2pap1} up to $\mathcal{O}(\partial^2)$ simplifies to
\begin{align}
 &\partial_{\mu}(u^{\mu}S^{\alpha\beta})= -2T^{\alpha\beta}_{\text{can}(1a)}-2\partial_{\mu}\Phi^{\mu\alpha\beta}_{\text{can(0)}}, \label{equ10pap1}
\end{align}
where for simplicity we have introduced the tensor $\Phi^{\mu\alpha\beta}_{\text{can(0)}}\equiv u^{[\alpha}S^{\beta]\mu}$ which is antisymmetric in the last two indices.
Thus, if we substitute equation~\eqref{equ10pap1} into Eq.~\eqref{equ9pap1}, we find
\begin{align} \partial_{\mu}\mathcal{S}^{\mu}_{\text{can}} &=T^{\alpha\beta}_{\text{can}(1s)}\partial_{\alpha}\beta_{\beta}+T^{\alpha\beta}_{\text{can}(1a)}\big[\partial_{\alpha}\beta_{\beta}+2\beta \omega_{\alpha\beta}\big] \nonumber \\ &+2\beta \omega_{\alpha\beta} \partial_{\mu}\Phi^{\mu\alpha\beta}_{\text{can(0)}} , \label{equ11pap1}
\end{align}
where we have decomposed $T^{\mu\nu}_{\text{can}(1)}$ into its symmetric and antisymmetric parts.
It should be noted that due to the totally antisymmetric structure of the canonical spin tensor, the conservation of the total angular momentum leads to mixing between the different components of the spin tensor, as can be clearly seen in Eq.~\eqref{equ10pap1}. This is a unique property of the canonical framework of spin hydrodynamics. 

To identify the transport coefficients in the canonical framework, we should be able to write Eq.~\eqref{equ11pap1} in terms of the dissipative currents. The symmetric and antisymmetric components of $T^{\alpha\beta}_{\text{can}(1)}$ can be decomposed in terms of the irreducible tensor basis as follows~\cite{Hattori:2019lfp,Fukushima:2020ucl}
\begin{align} & T^{\alpha\beta}_{\text{can}(1s)} = h^{\alpha}u^{\beta}+h^{\beta}u^{\alpha}+\tau^{\alpha\beta},\label{equ13pap1}\\ &T^{\alpha\beta}_{\text{can}(1a)} = q^{\alpha}u^{\beta}-q^{\beta}u^{\alpha}+\phi^{\alpha\beta},
\label{equ14pap1}
\end{align}
respectively. Here, $h^{\mu}$, $\tau^{\mu\nu}$, $q^{\mu}$ and $\phi^{\mu\nu}$ can be identified as different dissipative currents and satisfy the following conditions: $q \cdot u=0, h \cdot u=0, \tau^{\mu\nu}u_{\mu}=0, \phi^{\mu\nu}u_{\mu}=0, \tau^{\mu\nu}=\tau^{\nu\mu}, \phi^{\mu\nu}=-\phi^{\nu\mu}$. For the symmetric energy-momentum tensor, the heat flux $h^{\mu}$ and the viscous stress tensor $\tau^{\mu\nu}$ (which contains the shear and bulk viscosity terms) are the only dissipative fluxes. However, in the case of a non-vanishing antisymmetric part of the energy-momentum tensor, the new dissipative currents $q^{\mu}$ and $\phi^{\mu\nu}$ arise.

Analogously to $T^{\mu\nu}_{\text{can}(1a)}$ in Eq.~\eqref{equ14pap1}, we decompose the antisymmetric term $\partial_{\mu}\Phi^{\mu\alpha\beta}_{\text{can(0)}}$ and the spin chemical potential $\omega^{\alpha\beta}$ as 
\begin{align}
     \partial_{\mu}\Phi^{\mu\alpha\beta}_{\text{can(0)}}&=\delta q^{\alpha}u^{\beta}-\delta q^{\beta}u^{\alpha}+\delta\phi^{\alpha\beta},  \label{equ22pap1}\\
      \omega^{\alpha\beta}&=k^{\alpha}u^{\beta}-k^{\beta}u^{\alpha}+\lambda^{\alpha\beta},
    \label{equ26pap1}
\end{align}
where the components $\delta q$ and $k$ as well as $\delta \phi$ and $\lambda$ have the same properties as $q$ and $\phi$, respectively.
 
Moreover, using the decomposition $\partial_{\mu}\equiv \nabla_{\mu}+u_{\mu}D$, with $u^{\mu}\nabla_{\mu}=0, \nabla_{\mu}=\Delta^{\alpha}_{\mu}\partial_{\alpha}$ and $D=u^{\mu}\partial_{\mu}$, it can be shown that (for a detailed derivation see appendix~\ref{appenB}) 
\begin{align} \partial_{\mu}\mathcal{S}^{\mu}_{\text{can}}= &-\beta h^{\mu}\left(\beta \nabla_{\mu}T-Du_{\mu}\right)+\beta\pi^{\mu\nu}\sigma_{\mu\nu}+\beta\Pi \theta\nonumber\\ & -\beta q^{\mu}\left(\beta \nabla_{\mu}T+Du_{\mu}-4 \omega_{\mu\nu}u^{\nu}\right)\nonumber\\ & +\phi^{\mu\nu}\left(\Omega_{\mu\nu}+2\beta \Delta^{\alpha}_{~\mu}\Delta^{\beta}_{~\nu}\omega_{\alpha\beta}\right)\nonumber\\ & +2\beta\left[2k_{\alpha}\delta q^{\alpha}+\lambda_{\alpha\beta}\delta\phi^{\alpha\beta}\right]. \label{equ27pap1}
\end{align}
Here $\pi^{\mu\nu}$ is the traceless part of $\tau^{\mu\nu}$ and $\Pi$ is the trace of $\tau^{\mu\nu}$, i.e. $\tau^{\mu\nu}=\pi^{\mu\nu}+\Pi\Delta^{\mu\nu}$, $\sigma_{\mu\nu}\equiv\nabla_{(\mu} u_{\nu)}-\frac{1}{3}\theta\Delta_{\mu\nu}$ is the shear tensor with $\theta=\nabla^{\alpha}u_{\alpha}$, and $\Omega_{\mu\nu}\equiv\Delta^{\alpha}_{~\mu}\Delta^{\beta}_{~\nu}\partial_{[\alpha}\beta_{\beta]} = \beta \nabla_{[\mu} u_{\nu]}$ is the vorticity tensor.  

Imposing the second law of thermodynamics, i.e. $\partial_{\mu}\mathcal{S}^{\mu}_{\text{can}}\geq 0$ we obtain the conditions 
\begin{align} h^{\mu}&=-\kappa\left(Du^{\mu}-\beta\nabla^{\mu}T\right),\label{equ30pap1}\\ \pi^{\mu\nu}&=2\eta\sigma^{\mu\nu},\label{equ31pap1}\\ \Pi &= \zeta \theta,\label{equ32pap1}\\ q^{\mu}&=\lambda \left(\beta\nabla^{\mu}T+Du^{\mu}-4\omega^{\mu\nu}u_{\nu}\right),\label{equ33pap1}\\ \phi^{\mu\nu}&=\gamma\left(\Omega^{\mu\nu}+2\beta \Delta^{\mu}_{~\alpha}\Delta^{\nu}_{~\beta}\omega^{\alpha\beta}\right),\label{equ34pap1}
\end{align}
together with the constraint \footnote{One can in principle use on-shell conditions/ hydrodynamic equations to further simplify Eq.~\eqref{equ27pap1}. But even in that case one will end up with this additional constraint, see Appendix~\ref{appenE} for a detailed discussion. }
\begin{align}
2 k_{\alpha}\delta q^{\alpha}+ \lambda_{\alpha\beta}\delta\phi^{\alpha\beta}\geq 0.
\label{newcondition}
\end{align}
Here various transport coefficients can be identified with the conditions: $\kappa\geq 0$, $\eta\geq 0$, $\zeta\geq 0$, $\lambda \geq 0, \gamma\geq 0$. Note that $k^{\alpha}$, $\lambda^{\alpha\beta}$ as well as 
\begin{align} \delta q^{\alpha}=& \Delta^{\alpha}_{~\mu}u_{\nu}\partial_{\lambda}\Phi^{\lambda\mu\nu}_{\text{can(0)}},\label{equ23pap1}\\
 \delta\phi^{\alpha\beta}=& \Delta^{\alpha}_{~[\mu}\Delta^{\beta}_{~\nu]}\partial_{\lambda}\Phi^{\lambda\mu\nu}_{\text{can(0)}},\label{equ24pap1}
\end{align}
originate from completely different terms. Therefore, Eq.~\eqref{newcondition} is not satisfied for arbitrary initial conditions in general. The presence of such an additional condition implies that the canonical framework with the energy-momentum tensor and the spin tensor, as given in Eqs.~\eqref{equ3pap1} and \eqref{equ4pap1}, make the resulting hydrodynamic framework not a well-defined initial value problem for an arbitrary set of hydrodynamic variables, $T$, $u^{\mu}$ and $\omega^{\mu\nu}$ \footnote{In Ref.~\cite{Hu:2022azy}, a novel method has been introduced for the entropy current analysis to identify appropriate dissipative currents. Such a generalized study of the entropy current can lead to new cross terms in the dissipative currents and related transport coefficients that have not been reported in earlier research. It can be argued that even such a generalized analysis does not solve the problem associated with the entropy current analysis pointed out in the present analysis, where net baryon number density is zero. Please see appendix~\ref{appenE} for a detailed discussion.}. However, such a problem can be solved by the proper modification of the energy-momentum tensor, as we discuss in the next section. 

\section{Improved canonical framework}
\label{sec:3}
%
The constitutive relation of the energy-momentum tensor, as given in Eq.~\eqref{equ3pap1}, is not unique. 
In particular, one can add to it an additional totally divergent term, namely
\begin{align} & \widetilde{T}^{\mu\nu}_{\text{can}} = T^{\mu\nu}_{(0)}+T^{\mu\nu}_{\text{can}(1)} +\partial_{\lambda}\left (u^{\nu}S^{\mu\lambda}\right). \label{equ3pap2}
\end{align}
Since the additional term is $\mathcal{O}(\partial^1)$, it does not affect the conservation law of the energy-momentum tensor~\eqref{equ1pap1}, but it does affect the evolution of the spin tensor~\eqref{equ2pap1} (if we require the conservation of the total angular momentum). Moreover, in the presence of the spin density tensor $S^{\mu\nu}$, at the level of the first-order gradient expansion one can always add such a term to the energy-momentum tensor, as it is allowed by the Lorentz symmetry.

For the choice of the energy-momentum tensor in Eq.~\eqref{equ3pap2} and the spin tensor in Eq.~\eqref{equ4pap1}, the conservation of total angular momentum~\eqref{equ2pap1} implies
\begin{align} \partial_{\mu}S^{\mu\alpha\beta}_{\text{can}}&=-2T^{\alpha\beta}_{\text{can}(1a)}-\partial_{\mu}(u^{\beta}S^{\alpha\mu})+\partial_{\mu}(u^{\alpha}S^{\beta\mu})\nonumber\\ \implies \partial_{\mu}(u^{\mu}S^{\alpha\beta})&= -2T^{\alpha\beta}_{\text{can}(1a)}. \label{equ10pap2}
\end{align}
Similarly to the previous section, in this case we can also define the nonequilibrium entropy current up to first order terms in the gradient expansion as follows
\begin{align} \widetilde{\mathcal{S}}^{\mu}_{\text{can}}&=\widetilde{T}^{\mu\nu}_{\text{can}}\,\beta_{\nu}+p\,\beta^{\mu}-\omega_{\alpha\beta}S^{\alpha\beta}\beta^{\mu} +\mathcal{O}(\partial^2)\nonumber\\ & = \mathcal{S}^{\mu}_{(0)}+ \widetilde{T}^{\mu\nu}_{\text{can}(1)}\,\beta_{\nu} +\mathcal{O}(\partial^2). \label{equ6pap2}
\end{align}
However, compared to Eq.~\eqref{equ6pap1}, the first-order correction to the equilibrium energy-momentum tensor here includes an additional term $\partial_{\lambda}(u^{\nu}S^{\mu\lambda})$, see Eq.~\eqref{equ3pap2}. 

The divergence of the entropy current~\eqref{equ6pap2} can now be expressed as 
\begin{align} \partial_{\mu}\widetilde{\mathcal{S}}^{\mu}_{\text{can}} = \widetilde{T}^{\mu\nu}_{\text{can}(1)}\partial_{\mu}\beta_{\nu}-\beta\omega_{\alpha\beta}\partial_{\mu}(u^{\mu}S^{\alpha\beta})+\mathcal{O}(\partial^3),
\label{equ9pap2}
\end{align}
which, using Eqs.~\eqref{equ3pap2} and \eqref{equ10pap2}, can be further rewritten as
\begin{align} & \partial_{\mu}\widetilde{\mathcal{S}}^{\mu}_{\text{can}}= T^{\alpha\beta}_{\text{can}(1s)}\partial_{\alpha}\beta_{\beta}+T^{\alpha\beta}_{\text{can}(1a)}\left[\partial_{\alpha}\beta_{\beta}+2\beta \omega_{\alpha\beta}\right]\nonumber\\ &~~~~~~~~~~~~~~~~~~~~~~ +\partial_{\mu}\beta_{\nu}\partial_{\lambda}(u^{\nu}S^{\mu\lambda}).   \label{equ11pap2}
\end{align}
Note that the last term in the above equation equals a total derivative term $\partial_{\mu}\left[\beta_{\nu}\partial_{\lambda}(u^{\nu}S^{\mu\lambda})\right]$ which can be absorbed in the divergence of the entropy current and gives the following
\begin{align} & \partial_{\mu}\widetilde{\mathcal{S}}^{\prime\mu}_{\text{can}}=T^{\alpha\beta}_{\text{can}(1s)}\partial_{\alpha}\beta_{\beta}+T^{\alpha\beta}_{\text{can}(1a)}\left[\partial_{\alpha}\beta_{\beta}+2\beta \omega_{\alpha\beta}\right], \label{equ12pap2}
\end{align}
where
\begin{align} \widetilde{\mathcal{S}}^{\prime\mu}_{\text{can}} & =\widetilde{\mathcal{S}}^{\mu}_{\text{can}}-\beta_{\nu}\partial_{\lambda}\left(u^{\nu}S^{\mu\lambda}\right)\nonumber\\ & = \mathcal{S}^{\mu}_{(0)}+\beta_{\nu}T^{\mu\nu}_{\text{can}(1)} +\mathcal{O}(\partial^2). \label{equ13pap2}
\end{align}
It is interesting to note that the last result indicates that $\widetilde{\mathcal{S}}^{\prime\mu}_{\text{can}}
= {\mathcal{S}}^\mu_{\rm can}$, hence, our modification procedure does not change the entropy current of the original system (although it changes the dynamics of spin degrees of freedom to yield a positive entropy growth).

The right-hand side of Eq.~\eqref{equ12pap2} can be easily shown to be positive definite by properly identifying the dissipative currents. Indeed, after decomposing $T^{\mu\nu}_{\text{can}(1)}$ into the irreducible tensor basis, as discussed in the previous section, we obtain
\begin{align} \partial_{\mu}\widetilde{\mathcal{S}}^{\prime\mu}_{\text{can}} = & -\beta h^{\mu}\left(\beta \nabla_{\mu}T-Du_{\mu}\right)+\beta\pi^{\mu\nu}\sigma_{\mu\nu}+\beta\Pi \theta\nonumber\\ & -\beta q^{\mu}\left (\beta \nabla_{\mu}T+Du_{\mu}-4 \omega_{\mu\nu}u^{\nu}\right)\nonumber\\ & +\phi^{\mu\nu}\left(\Omega_{\mu\nu}+2\beta \Delta^{\alpha}_{~\mu}\Delta^{\beta}_{~\nu}\omega_{\alpha\beta}\right). \label{equ22pap2}
\end{align}
Note that, unlike Eq.~\eqref{equ27pap1}, Eq.~\eqref{equ22pap2} is free of the potentially problematic terms due to the use of an improved energy-momentum tensor from~\eqref{equ3pap2}. If we impose the second law of thermodynamics, we can identify the dissipative currents as in Eqs.~(\ref{equ30pap1})-(\ref{equ34pap1}), but without the constraint~\eqref{newcondition}.
Without the additional condition~\eqref{newcondition}, the improved version of the canonical framework is well defined and the dissipative currents can be uniquely expressed by the hydrodynamic variables, i.e. $T$, $u^{\mu}$ and $\omega^{\mu\nu}$. 

Some remarks on Eqs.~(\ref{equ30pap1})-(\ref{equ34pap1}) are in order here. Note that the expressions of $h^{\mu}$, $\pi^{\mu\nu}$, and $\Pi$ in terms of the transport coefficients, i.e. heat conductivity $(\kappa)$, shear viscosity ($\eta$), and bulk viscosity ($\zeta$), are expected from relativistic Navier-Stokes hydrodynamics. The new coefficients $\lambda$ and $\gamma$ appear in the dissipative formulation of hydrodynamic with spin and were also found earlier~\cite{Hattori:2019lfp,Fukushima:2020ucl}. It should be emphasized that the transport coefficients $\kappa$, $\eta$, $\zeta$, $\lambda$ and $\gamma$ enter directly into the expression of the energy-momentum tensor and are expected to play an important role in the equilibration of the system and in the fluctuation-dissipation relations. 

\section{Transition from improved canonical to phenomenological formulation}
\label{sec:4}
%
In the previous section, we discussed in detail the canonical framework of dissipative spin hydrodynamics with dissipative currents uniquely determined by the positivity of the divergence of the entropy current. The canonical framework uses the energy-momentum tensor, which contains both symmetric and antisymmetric parts, and the canonical spin tensor is totally antisymmetric. The latter property follows directly from a microscopic theory~\cite{Peng:2021ago}, but does not provide a natural identification of the tensor $S^{\mu\nu}$ in Eq.~\eqref{equ4pap1} as the spin density~\footnote{If one additionally uses the Frenkel condition $S^{\mu\nu} u_\nu = 0$, then $S^{\mu\nu}$ can indeed be identified as the spin density, which in this case follows from the definition of the canonical spin tensor.}. Therefore, many formulations of spin hydrodynamics use a simpler form of the spin tensor with $S^{\mu\alpha\beta}_{\text{ph}(0)}=u^{\mu}S^{\alpha\beta}$. In this paper we refer to such formulations as to phenomenological ones and use the subscript ''ph'' in this context.

It is well known that one can always generate a new set of energy-momentum and spin tensors from a given set of such tensors. This is achieved by performing the so-called pseudogauge transformation \cite{Florkowski:2018fap,HEHL197655,Speranza:2020ilk,BELINFANTE1939887,BELINFANTE1940449,Rosenfeld1940}. In this section we show that it is possible to construct explicitly a pseudogauge leading directly from the (improved) canonical formulation given by the set of tensors~\eqref{equ4pap1} and \eqref{equ3pap2} to the phenomenological one. The starting point is 
\begin{align} & T^{\mu\nu}_{\text{ph}}=\widetilde{T}^{\mu\nu}_{\text{can}}+\frac{1}{2}\partial_{\lambda}\left(\Sigma^{\lambda\mu\nu}-\Sigma^{\mu\lambda\nu}-\Sigma^{\nu\lambda\mu}\right),\label{equ28pap2}\\ & S^{\lambda\mu\nu}_{\text{ph}}=S^{\lambda\mu\nu}_{\text{can}}-\Sigma^{\lambda\mu\nu},\label{equ29pap2}
\end{align}
where the pseudogauge potential is of the form
\begin{align} \Sigma^{\lambda\mu\nu}=2\Phi^{\lambda\mu\nu}_{\text{can(0)}}+\Sigma_{(1)}^{\lambda\mu\nu},
\label{equ30pap2}
\end{align}
where we introduced $\Sigma_{(1)}^{\lambda\mu\nu}$ to include all higher order correction terms in the spin current. Note that here both $\Sigma^{\lambda\mu\nu}$ and $\Sigma^{\lambda\mu\nu}_{(1)}$ are antisymmetric only in the last two indices.

Using the pseudogauge transformation we obtain (up to terms of order $\mathcal{O}(\partial)$)
\begin{align} S^{\lambda\mu\nu}_{\text{ph}} & = u^{\lambda}S^{\mu\nu}+S_{\text{ph(1)}}^{\lambda\mu\nu} \label{equ31pap2}
\end{align}
and
\begin{align} T^{\mu\nu}_{\text{ph}} & = \widetilde{T}^{\mu\nu}_{\text{can}}-\partial_{\lambda}(u^{\nu}S^{\mu\lambda})\nonumber\\ & = T^{\mu\nu}_{(0)}+T^{\mu\nu}_{\text{can}(1)}\nonumber\\ & =T^{\mu\nu}_{(0)}+T^{\mu\nu}_{\text{ph}(1)}, \label{equ32pap2}
\end{align}
where $S_{\text{ph(1)}}^{\lambda\mu\nu}=S^{\lambda\mu\nu}_{\text{can}(1)}-\Sigma_{(1)}^{\lambda\mu\nu}$ is first order in gradient expansion.

Having determined the energy-momentum tensor and the spin tensor for the phenomenological framework, we can study the nonequilibrium entropy current, which can be expressed as follows
\begin{align} \mathcal{S}^{\mu}_{\text{ph}} & = \mathcal{S}^{\mu}_{(0)}+\beta_{\nu}T^{\mu\nu}_{\text{ph}(1)}+\mathcal{O}(\partial^2). \label{equ33pap2}
\end{align}
Using the conservation equation of the phenomenological energy-momentum tensor, the divergence of the phenomenological entropy current can be simplified as
\begin{align} \partial_{\mu}\mathcal{S}^{\mu}_{\text{ph}} = T^{\mu\nu}_{\text{ph}(1)}\partial_{\mu}\beta_{\nu}-\beta\omega_{\alpha\beta}\partial_{\mu}(u^{\mu}S^{\alpha\beta})+\mathcal{O}(\partial^3).
\label{equ34pap2}
\end{align}
From Eqs.~\eqref{equ9pap2} and \eqref{equ34pap2}, we can see that the divergence of the canonical entropy current and the divergence of the phenomenological entropy current are not the same. However, $\partial_{\mu}\widetilde{\mathcal{S}}^{\mu}_{\text{can}}$ and $\partial_{\mu}\mathcal{S}^{\mu}_{\text{ph}}$ differ only by a total divergence term, see Eq.~\eqref{equ11pap2}. Using the expression of $T^{\mu\nu}_{\text{ph}(1)}$ \eqref{equ32pap2} in Eq.~\eqref{equ34pap2} we find,
\begin{align} \partial_{\mu}\mathcal{S}^{\mu}_{\text{ph}} & =\partial_{\mu}\beta_{\nu}T^{\mu\nu}_{\text{can}(1s)}+\bigg[\partial_{\alpha}\beta_{\beta}+2\beta \omega_{\alpha\beta}\bigg]T^{\alpha\beta}_{\text{can}(1a)}. \label{equ35pap2}
\end{align}
In the last equation, we used conservation of total angular momentum for the phenomenological currents. From the equations~\eqref{equ13pap2} and \eqref{equ33pap2}, we see that $\widetilde{\mathcal{S}}^{\prime\mu}_{\text{can}}
= {\mathcal{S}}^\mu_{\rm can}
=\mathcal{S}^{\mu}_{\text{ph}}$. Moreover, from Eqs.~\eqref{equ12pap2} and~\eqref{equ35pap2}, we can conclude that the entropy constraints are the same in the canonical and phenomenological formalisms. Using the tensor decomposition of $T^{\alpha\beta}_{\text{can}(1s)}$ and $T^{\alpha\beta}_{\text{can}(1a)}$ together with the condition $\partial_{\mu}\mathcal{S}^{\mu}_{\text{ph}}\geq 0$ we get back the dissipative currents as given in Eqs.~(\ref{equ30pap1})-(\ref{equ34pap1}). Note that the expressions for the dissipative currents as given in Eqs.~(\ref{equ30pap1})-(\ref{equ34pap1}) correspond exactly to the dissipative currents obtained in Ref.~\cite{Hattori:2019lfp}. Thus, we conclude that starting from the improved canonical framework, it is indeed possible to recover the phenomenological formalism of first-order dissipative spin hydrodynamics.

\section{Improved canonical framework starting from the phenomenological formalism}
\label{sec:5}
%
Till now we have derived the dissipative spin hydrodynamics framework starting from the canonical formalism with a general structure of the energy-momentum tensor and the totally antisymmetric spin tensor. In such a framework we uniquely determined the dissipative currents using the entropy current analysis. Furthermore, using the pseudogauge transformation we have obtained the phenomenological energy-momentum and spin tensor and also studied the dissipative spin hydrodynamics in this framework. The energy-momentum tensor in the phenomenological as well as in the canonical framework contains symmetric and antisymmetric parts. However, contrary to the phenomenological spin tensor, in the canonical framework the spin tensor is always totally antisymmetric. 

Note that to achieve a well defined description of dissipative spin hydrodynamics we have improved the energy-momentum tensor adding to the standard dissipative part a totally divergent term. We have argued that in the presence of a spin density tensor $S^{\mu\nu}$ such a term is very natural. In this section we will explicitly show that such a term automatically arises if one considers the pseudogauge transformation that starts from the phenomenological forms. We also argue that additional term which originates from the pseudogauge transformation also modifies various dissipative currents. Appearance of the pseudogauge corrections to various dissipative currents is straightforward to understand: as the pseudogauge transformation enters into the expression of the energy-momentum tensor at $\mathcal{O}(\partial)$, it effectively introduces new dissipative terms in the original energy-momentum tensor. To put the above statement on the firm mathematical ground here we will first discuss the framework of dissipative spin hydrodynamics using the phenomenological energy-momentum tensor and spin tensor as introduced in Ref.~\cite{Hattori:2019lfp}. After that, we will derive the canonical framework using pseudogauge transformation. 

In the phenomenological framework we start with the constitutive relations for $T^{\mu\nu}_{\text{ph}}$ and $S^{\mu\alpha\beta}_{\text{ph}}$ up to first order in the gradient expansion \footnote{Note that the calculations discussed in this section are completely independent of the previous discussions. However, without introducing new notations for the energy-momentum tensor and the spin tensor we use the same as introduced earlier in the manuscript.}~\cite{Hattori:2019lfp},
\begin{align}
    & T^{\mu\nu}_{\text{ph}} = T^{\mu\nu}_{(0)}+T^{\mu\nu}_{\text{ph}(1)}, \label{equ36pap2}\\
    & S^{\mu\alpha\beta}_{\text{ph}}=u^{\mu}S^{\alpha\beta}+S^{\mu\alpha\beta}_{\text{ph}(1)}. \label{equ37pap2}
\end{align}  
In general, $T^{\mu\nu}_{\text{ph}(1)}$ contains symmetric as well as antisymmetric components, while $S^{\mu\alpha\beta}_{\text{ph}(1)}$ is a tensor antisymmetric in last two indices only. This also implies that the spin tensor in the phenomenological framework is only antisymmetric in last two indices. 

Analogous to the previous sections, we define the nonequilibrium entropy current up to the first-order term in the gradient expansion as
\begin{align} \mathcal{S}^{\mu}_{\text{ph}}& =\beta_{\nu}T^{\mu\nu}_{\text{ph}}+ p \beta^{\mu}- \beta\omega_{\alpha\beta}S^{\alpha\beta}u^{\mu}+\mathcal{O}(\partial^2). \label{equ38pap2}
\end{align}

The divergence of Eq.~\eqref{equ38pap2} can then be expressed in the form
\begin{align} \partial_{\mu}\mathcal{S}^{\mu}_{\text{ph}}  & = T^{\mu\nu}_{\text{ph}(1s)}\partial_{\mu}\beta_{\nu}+T^{\mu\nu}_{\text{ph}(1a)}(\partial_{\mu}\beta_{\nu}+2\beta\omega_{\mu\nu})+\mathcal{O}(\partial^3). \label{equ41pap2}
\end{align}
Decomposing the symmetric and antisymmetric components of the phenomenological energy-momentum tensor as
$T^{\mu\nu}_{\text{ph}(1s)} = \overline{h}^{\mu}u^{\nu}+\overline{h}^{\nu}u^{\mu}+\overline{\tau}^{\mu\nu}$ and $T^{\mu\nu}_{\text{ph}(1a)} = \overline{q}^{\mu}u^{\nu}-\overline{q}^{\nu}u^{\mu}+\overline{\phi}^{\mu\nu}$, we obtain the following constraint on the divergence of the phenomenological entropy current
\begin{align} \partial_{\mu}\mathcal{S}^{\mu}_{\text{ph}} = & -\beta \overline{h}^{\mu}\bigg(\beta \nabla_{\mu}T-Du_{\mu}\bigg)+\beta \overline{\pi}^{\mu\nu}\sigma_{\mu\nu}+\beta\overline{\Pi}\theta\nonumber\\ & -\beta\overline{q}^{\mu}\bigg(\beta\nabla_{\mu}T+Du_{\mu}-4u^{\nu}\omega_{\mu\nu}\bigg)\nonumber\\ & +\overline{\phi}^{\mu\nu}\left(\Omega_{\mu\nu}+2\beta \Delta^{\alpha}_{\mu}\Delta^{\beta}_{\nu}\omega_{\alpha\beta}\right) \geq 0 \label{equ49pap2}
\end{align}
with dissipative currents $\overline{h}, \overline{\pi}, \overline{\Pi}, \overline{q}$ and $\overline{\phi}$ satisfying relations similar to those shown in Eqs.~(\ref{equ30pap1})--(\ref{equ34pap1}).\footnote{If the kinetic coefficients are independent of the pseudogauge, which is reasonable to assume, the tensors $\overline{h}, \overline{\pi}, \overline{\Pi}, \overline{q}$ and $\overline{\phi}$ should be equal to their counterparts without a bar. }

Let us now discuss the improved canonical framework that results from applying the pseudogauge transformation to the phenomenological framework. In this case, using the pseudogauge transformation as
\begin{align} & T^{\mu\nu} = T^{\mu\nu}_{\text{ph}}+\frac{1}{2}\partial_{\lambda}\bigg(\Psi^{\lambda\mu\nu}-\Psi^{\mu\lambda\nu}-\Psi^{\nu\lambda\mu}\bigg),\label{equ55pap2}\\ & S^{\mu\alpha\beta}=S^{\mu\alpha\beta}_{\text{ph}}-\Psi^{\mu\alpha\beta},\label{equ56pap2}
\end{align} 
and choosing the pseudogauge potential $\Psi^{\lambda\mu\nu}$ in the form
\begin{align}
\Psi^{\mu\alpha\beta}=S_{\text{ph}}^{\alpha\mu\beta}-S_{\text{ph}}^{\beta\mu\alpha},
\label{equ57pap2}
\end{align}
we obtain the following constitutive relations
\begin{align} T^{\mu\nu} =\widetilde{T}^{\mu\nu}_{\text{can}}&=T^{\mu\nu}_{\text{ph}}+\delta T^{\mu\nu}_\text{ph(1)}\label{equ59pap2},\\ S^{\mu\alpha\beta}=S_{\text{can}}^{\mu\alpha\beta}&=S_{\text{ph}}^{\mu\alpha\beta}+S_{\text{ph}}^{\beta\mu\alpha}+S_{\text{ph}}^{\alpha\beta\mu}, \label{equ58pap2}
\end{align}
where $S_{\text{can}}^{\mu\alpha\beta}$ is totally antisymmetric and we have used in Eq.~\eqref{equ59pap2} the expression of $S^{\nu\mu\lambda}_{\text{ph}}$ from Eq.~\eqref{equ37pap2}.

It should be emphasized that the term $\delta T^{\mu\nu}_\text{ph(1)}=\partial_{\lambda}\left(u^{\nu}S^{\mu\lambda}\right)$ was introduced in the previous section in a heuristic way to obtain a well-defined description within the canonical formalism. Interestingly, such a term is obtained here in a natural way applying the pseudogauge transformation to the phenomenological framework. 

One can observe that the additional term $\delta T^{\mu\nu}_\text{ph(1)}$ gives rise to the first-order corrections for various components of the initial energy-momentum tensor, which are additional to those arising from $T^{\mu\nu}_{\text{ph}(1)}$. Thus, the resulting energy-momentum tensor can be written in the following form
\begin{align} \widetilde{T}^{\mu\nu}_{\text{can}}&=T^{\mu\nu}_{(0)}+T^{\mu\nu}_{\text{ph}(1)}+ \delta T^{\mu\nu}_\text{ph(1)}, \\ & = \widetilde{\varepsilon} u^{\mu}u^{\nu}-p\Delta^{\mu\nu}+\widetilde{h}^{\mu}u^{\nu}+\widetilde{h}^{\nu}u^{\mu}+\widetilde{\tau}^{\mu\nu}\nonumber\\ & ~~~~~~~~~~~~~~~~~~~~~~~~+\widetilde{q}^{\mu}u^{\nu}-\widetilde{q}^{\nu}u^{\mu}+\widetilde{\phi}^{\mu\nu},
\label{newTcan}
\end{align}
where
\begin{align} & \widetilde{\varepsilon}=\varepsilon+\delta\overline{\varepsilon},\label{c1}\\ & \widetilde{h}^{\mu}=\overline{h}^{\mu}+\delta\overline{h}^{\mu},\label{c2}\\ & \widetilde{\tau}^{\mu\nu}=\overline{\tau}^{\mu\nu}+\delta\overline{\tau}^{\mu\nu},\label{c3}\\ & \widetilde{q}^{\mu}=\overline{q}^{\mu}+\delta\overline{q}^{\mu},\label{c4}\\ & \widetilde{\phi}^{\mu\nu}=\overline{\phi}^{\mu\nu}+\delta\overline{\phi}^{\mu\nu}.\label{c5}
\end{align}
For the details of the decomposition in Eq.~\eqref{newTcan} and the definitions of the additional $\delta$ contributions in  Eqs.~\eqref{c1}-\eqref{c5} see appendix~\ref{appenC}. 

As mentioned above, the nonequilibrium entropy current can be defined as
\begin{align} \widetilde{S}^{\mu}_{\text{can}}& =\beta_{\nu}\widetilde{T}^{\mu\nu}_{\text{can}}+ p\beta^{\mu}-\beta^\mu \omega_{\alpha\beta}S^{\alpha\beta} +\mathcal{O}(\partial^2)\nonumber\\ & = {\cal S}_{(0)}^{\mu}+\beta_{\nu}\widetilde{T}^{\mu\nu}_{\text{can}(1)} +\mathcal{O}(\partial^2). \label{equ80pap2}
\end{align}
Using the conservation of canonical energy-momentum tensor, the divergence of the canonical entropy current can be written as follows
\begin{align} \partial_{\mu}\widetilde{S}^{\mu}_{\text{can}}& = \partial_{\mu}\beta_{\nu}\widetilde{T}^{\mu\nu}_{\text{can}(1)}-\beta \omega_{\alpha\beta}\partial_{\mu}(u^{\mu}S^{\alpha\beta})\nonumber+\mathcal{O}(\partial^3)\\ & = \partial_{\mu}\beta_{\nu} T^{\mu\nu}_{\text{ph}(1)}+\partial_{\mu}\bigg[\beta_{\nu}\partial_{\lambda}(u^{\nu}S^{\mu\lambda}) \bigg]\nonumber\\ & ~~~~~~~~~~-\beta \omega_{\alpha\beta}\partial_{\mu}(u^{\mu}S^{\alpha\beta})+\mathcal{O}(\partial^3). \label{equ81pap2}
\end{align}
Absorbing that total derivative term into the definition of the entropy current, we get 
\begin{align}
& \partial_{\mu}\mathcal{S}^{\mu}_{\text{ph}}= \partial_{\mu}\beta_{\nu} T^{\mu\nu}_{\text{ph}(1)} 
-\beta \omega_{\alpha\beta}\partial_{\mu}(u^{\mu}S^{\alpha\beta})+\mathcal{O}(\partial^3)
\label{equ82pap2}
\end{align}
where $\mathcal{S}^{\mu}_{\text{ph}} = \widetilde{S}^{\mu}_{\text{can}}-\beta_\nu\partial_{\lambda}(u^{\nu}S^{\mu\lambda}) = \mathcal{S}^{\mu}_{\text{can}}$.
Thus, we conclude that the entropy current constraints in these two frameworks are equivalent. 

\section{Summary and conclusions}
\label{sec:6}

We formulate the first-order dissipative spin hydrodynamics for the canonical framework with a totally antisymmetric spin tensor. We argue that the naive use of a general energy-momentum tensor containing both a symmetric and an antisymmetric part, together with a totally antisymmetric spin tensor, is in conflict with the principle of entropy production. We show that this problem can be solved by an improved form of the canonical energy-momentum tensor, obtained by modifying the canonical energy-momentum tensor with a suitable totally divergent term that does not affect the conservation of the energy-momentum tensor. The origin of such a term in the improved canonical framework can also be justified using the concept of the pseudogauge transformation starting from the phenomenological formalism. Using such an improved form, we show that the framework of dissipative spin hydrodynamics is well defined within the canonical framework. We argue that one can always recover the phenomenological framework of spin hydrodynamics from such an improved canonical framework using an appropriate pseudogauge transformation, and vice versa. Therefore, the improved canonical framework and the phenomenological framework of first-order dissipative spin hydrodynamics are equivalent since they are connected by a suitable pseudogauge transformation. Although a pseudogauge transformation can lead to modified dissipative currents, entropy current analysis shows that the entropy production condition for a dissipative system remains unchanged in these two frameworks, i.e., the entropy current constraint is independent of a pseudogauge transformation.


\medskip
{\it Acknowledgements.} 
This work was supported in part by the Polish National Science Centre Grants No. 2022/47/B/ST2/01372, No. 2018/30/E/ST2/00432, and No. 2020/39/D/ST2/02054. A.D. would like to thank Shi Pu for some important discussions and clarifications on the formulation of dissipative spin hydrodynamics.
\appendix
%
%
%
   %
    %
%
%
\section{DERIVATION OF EQ.~\eqref{equ27pap1}}
\label{appenB}
%
Here we present the detailed derivation of Eq.~\eqref{equ27pap1}. We start with Eq.~\eqref{equ11pap1} which using the decomposition of symmetric and antisymmetric components of $T^{\mu\nu}_{\text{can}(1)}$ in terms of the dissipative currents as given in Eqs.~\eqref{equ13pap1} and \eqref{equ14pap1} can be rewritten as
\begin{align}
    \partial_{\mu}\mathcal{S}^{\mu}_{\text{can}}&= 2\beta\omega_{\alpha\beta}\left[T^{\alpha\beta}_{\text{can}(1a)}+\partial_{\mu}\Phi^{\mu\alpha\beta}_{\text{can(0)}}\right]\nonumber\\
    & ~~~~+h^{\mu}u^{\nu}(\partial_{\mu}\beta_{\nu}+\partial_{\nu}\beta_{\mu})+\tau^{\mu\nu}\partial_{\mu}\beta_{\nu}\nonumber\\
    & ~~~~+q^{\mu}u^{\nu}(\partial_{\mu}\beta_{\nu}-\partial_{\nu}\beta_{\mu})+\phi^{\mu\nu}\partial_{\mu}\beta_{\nu}.
    \label{appdiv}
\end{align}

Using the decomposition $\partial_{\mu}\equiv \nabla_{\mu}+u_{\mu}D$, with $u^{\mu}\nabla_{\mu}=0, \nabla_{\mu}=\Delta^{\alpha}_{\mu}\partial_{\alpha}$, and $D=u^{\mu}\partial_{\mu}$, it can be shown that
\begin{align}
    \partial_{\mu}\beta_{\nu}=u_{\nu}\nabla_{\mu}\beta+u_{\nu}u_{\mu}D\beta+\beta\nabla_{\mu}u_{\nu}+\beta u_{\mu}Du_{\nu}. 
    \label{equ16pap1}
\end{align}
This implies
\begin{align}
     h^{\mu}u^{\nu}\left(\partial_{\mu}\beta_{\nu}+\partial_{\nu}\beta_{\mu}\right)
 = & -\beta h^{\mu}\left(\beta \nabla_{\mu}T-Du_{\mu}\right),
\label{equ17pap1}
\end{align}
and
\begin{align}
    q^{\mu}u^{\nu}\left(\partial_{\mu}\beta_{\nu}-\partial_{\nu}\beta_{\mu}\right) =-\beta q^{\mu}\left(\beta\nabla_{\mu}T+Du_{\mu}\right). 
    \label{equ18pap1}
\end{align}
The symmetric tensor $\tau^{\mu\nu}$ can be decomposed into trace and traceless part, $\tau^{\mu\nu}=\pi^{\mu\nu}+\Pi\Delta^{\mu\nu}$. Using Eq.~\eqref{equ16pap1} and the fact that $\tau^{\mu\nu}$ is orthogonal to the fluid four velocity we obtain
\begin{align}
    & \tau^{\mu\nu}\partial_{\mu}\beta_{\nu}=\tau^{\mu\nu}\beta\nabla_{\mu}u_{\nu}=\beta\left(\pi^{\mu\nu}+\Pi \Delta^{\mu\nu}\right)\nabla_{\mu}u_{\nu}\nonumber\\
    & = \beta\pi^{\mu\nu}\bigg[\frac{1}{2}\left(\nabla_{\mu}u_{\nu}+\nabla_{\nu}u_{\mu}\right)-\frac{1}{3}\Delta_{\mu\nu}\nabla^{\alpha}u_{\alpha}+\frac{1}{3}\Delta_{\mu\nu}\nabla^{\alpha}u_{\alpha}\bigg]\nonumber\\
    & ~~~~~~~~~~~~~~~~~~~+\beta \Pi \Delta^{\mu\nu}\nabla_{\mu}u_{\nu}\nonumber\\
    & = \beta\pi^{\mu\nu}\bigg[\nabla_{(\mu}u_{\nu)} -\frac{1}{3}\theta\Delta_{\mu\nu}\bigg]+\beta\Pi  \theta\nonumber\\
    & = \beta \pi^{\mu\nu}\sigma_{\mu\nu}+\beta\Pi  \theta,
    \label{equ19pap1}
\end{align}
where $\theta=\nabla^{\alpha}u_{\alpha}$.
In the third line of the above equation we have used the property that $\pi^{\mu\nu}$ is traceless, i.e. $\pi^{\mu}_{~\mu}=0$ and $\pi^{\mu\nu}u_{\nu}=0=\pi^{\mu\nu}u_{\mu}$. It is also important to note that $\sigma_{\mu\nu}=\nabla_{(\mu}u_{\nu)} -\frac{1}{3}\theta\Delta_{\mu\nu}$ is traceless, i.e. $\sigma^{\mu}_{~\mu}=0$, and  orthogonal to the fluid four velocity, i.e. $\sigma^{\mu\nu}u_{\mu}=0=\sigma^{\mu\nu}u_{\nu}$. 

Moreover it can be shown that
\begin{align}
    &\phi^{\mu\nu} \partial_{\mu}\beta_{\nu}=\frac{1}{2}\phi^{\mu\nu}\left(\partial_{\mu}\beta_{\nu}-\partial_{\nu}\beta_{\mu}\right)= \phi^{\mu\nu} \partial_{[\mu}\beta_{\nu]} \nonumber\\
    & = \beta \phi^{\mu\nu}\nabla_{[\mu}u_{\nu]}\nonumber\\
    & = \frac{\beta}{2}\phi^{\mu\nu}\left(\Delta^{\alpha}_{\mu}\partial_{\alpha}u_{\nu}-\Delta^{\alpha}_{\nu}\partial_{\alpha}u_{\mu}\right)\nonumber\\
    & = \phi^{\mu\nu}\Delta^{\alpha}_{~\mu}\Delta^{\beta}_{~\nu}\partial_{[\alpha}\left(\beta u_{\beta]}\right)=\phi^{\mu\nu}\Omega_{\mu\nu},
\label{equ20pap1}
\end{align}
where we defined $\Omega_{\mu\nu}\equiv\Delta^{\alpha}_{~\mu}\Delta^{\beta}_{~\nu}\partial_{[\alpha}\beta_{\beta]} = \beta \nabla_{[\mu} u_{\nu]}$.

Using Eqs.~\eqref{equ17pap1}-\eqref{equ20pap1}, the divergence of the entropy current as given in Eq.~\eqref{appdiv} can be expressed as
 \begin{align}
    \partial_{\mu}\mathcal{S}^{\mu}_{\text{can}}& = 2\beta\omega_{\alpha\beta} \partial_{\mu}\Phi^{\mu\alpha\beta}_{\text{can(0)}} \nonumber\\
    & ~-\beta h^{\mu}\bigg(\beta \nabla_{\mu}T-Du_{\mu}\bigg)+\beta \pi^{\mu\nu}\sigma_{\mu\nu}+\beta\Pi \partial^{\alpha}u_{\alpha}\nonumber\\
    & -\beta q^{\mu}\left(\beta \nabla_{\mu}T+Du_{\mu}-4 \omega_{\mu\nu}u^{\nu}\right)\nonumber\\
     & +\phi^{\mu\nu}\left(\Omega_{\mu\nu}+2\beta \Delta^{\alpha}_{~\mu}\Delta^{\beta}_{~\nu}\omega_{\alpha\beta}\right) 
    \label{}
\end{align}
Using in the above equation decompositions~\eqref{equ22pap1}-\eqref{equ26pap1} we obtain Eq.~\eqref{equ27pap1}.
%
\section{ESTABLISHING THE IMPERATIVENESS
OF INDEPENDENT TREATMENT FOR DIFFERENT DISSIPATIVE CURRENTS}
\label{appenE}
In the work presented in Ref.~\cite{Hu:2022azy}, a novel approach has been introduced for analyzing the entropy current in order to correctly identify dissipative terms. This comprehensive investigation of the entropy current has the potential to reveal previously unreported cross terms in the dissipative currents as well as uncover new transport coefficients. However, here we argue that at zero chemical potential, different dissipative currents should indeed be considered independently without any cross-effects. Moreover, we also show that hydrodynamic equations and associated hydrodynamic gradient expansion do not affect our conclusion. For a clear demonstration, let us start with the following part of the entropy production equation (associated with the dissipative currents, which are four-vectors, i.e., $h^{\mu}$, and $q^{\mu}$ as given in Eq.~\eqref{equ27pap1}),
\begin{align}
\partial_{\mu}\mathcal{S}^{\mu}_{\text{can} (h,q)} & =-\beta h^{\mu}\left(\beta\nabla_{\mu}T-Du_{\mu}\right)\nonumber\\
& -\beta q^{\mu}\left(\beta \nabla_{\mu}T+Du_{\mu}-4\omega_{\mu\nu}u^{\nu}\right).  
\label{equappenE1}
\end{align}
Following the arguments presented in Ref.~\cite{Hu:2022azy} in general one can write, 
\begin{align}
    h^{\mu}& =a\left(\beta\nabla^{\mu}T-Du^{\mu}\right)+b\left(\beta \nabla^{\mu}T+Du^{\mu}-4\omega^{\mu\nu}u_{\nu}\right)\nonumber\\
    &\equiv aX^{\mu}+bY^{\mu}, 
    \label{equappenE2}
\end{align}
and, 
\begin{align}
    q^{\mu} & =c\left(\beta\nabla^{\mu}T-Du^{\mu}\right)+d\left(\beta \nabla^{\mu}T+Du^{\mu}-4\omega^{\mu\nu}u_{\nu}\right)\nonumber\\
    &\equiv cX^{\mu}+dY^{\mu}. 
\label{equappenE3}
\end{align}
Therefore, plugging Eqs.~\eqref{equappenE2} and \eqref{equappenE3} back into Eq.~\eqref{equappenE1} we can write, 
\begin{align}
\partial_{\mu}\mathcal{S}^{\mu}_{\text{can} (h,q)} & =-\beta X_{\mu}\left(aX^{\mu}+bY^{\mu}\right) -\beta Y_{\mu}\left(cX^{\mu}+dY^{\mu}\right)\nonumber\\
& = -a\beta X^{\mu}X_{\mu}-\beta d Y^{\mu}Y_{\mu}-\beta (b+c) X^{\mu}Y_{\mu}.
\end{align}
Note that $X^{\mu}u_{\mu}=0=Y^{\mu}u_{\mu}$. Hence $X^{\mu}$ and $Y^{\mu}$ are space-like and we can write, $X^{\mu}X_{\mu}=-X^{2}$, $Y^{\mu}Y_{\mu}=-Y^{2}$, with $X^2\geq0$, $Y^2\geq0$. However, we cannot say anything about the positivity of the product $X^{\mu}Y_{\mu}$. Nevertheless, we can always choose a frame where, $X^{\mu}\equiv (0,\vec{X})$, and $Y^{\mu}\equiv (0,\vec{Y})$, which implies $X^2=\vec{X}\cdot \vec{X}$, $Y^2=\vec{Y}\cdot \vec{Y}$, and $X^{\mu}Y_{\mu}=-\vec{X}\cdot\vec{Y}$. We can further simplify by representing $\vec{X}\equiv(X^1,X^2,X^3)$ and $\vec{Y}\equiv(Y^1,Y^2,Y^3)$,
\begin{align}
    \partial_{\mu}\mathcal{S}^{\mu}_{\text{can} (h,q)} & =a\beta \left[(X^1)^2+(X^2)^2+(X^3)^2\right]\nonumber\\
    & +\beta d \left[(Y^1)^2+(Y^2)^2+(Y^3)^2\right]\nonumber\\
    & +\beta(b+c) \left[X^1Y^1+X^2Y^2+X^3Y^3\right]. 
\end{align}
Hence, the condition for the entropy production implies, 
\begin{align}
    & a\beta \left[(X^1)^2+(X^2)^2+(X^3)^2\right]\nonumber\\
    & +\beta d \left[(Y^1)^2+(Y^2)^2+(Y^3)^2\right]\nonumber\\
    & +\beta(b+c) \left[X^1Y^1+X^2Y^2+X^3Y^3\right]\geq 0.
    \label{equappenE6}
\end{align}
The above equation leads to the following conditions, 
\begin{align}
    a\beta\geq 0, \quad \beta d\geq 0, \quad 4ad\beta^2\geq (b+c)^2\beta^2.
\end{align}
The above discussion fixes the sign of various coefficients ($a,b,c,d$) in the dissipative currents. Interestingly the above discussion does not imply that $b=0$, or $c=0$. Note that only for $b=0$ and $c=0$ we get back our results (Eqs.~\eqref{equ30pap1}, \eqref{equ33pap1}), i.e.,   
\begin{align}
     & h^{\mu}=a X^{\mu}=a \left(\beta\nabla^{\mu}T-Du^{\mu}\right), \quad a\geq 0.\label{equappenE8}\\ 
     & q^{\mu}= d \left(\beta \nabla^{\mu}T+Du^{\mu}-4\omega^{\mu\nu}u_{\nu}\right), d\geq 0\label{equappenE9}. 
\end{align}
However, when $b\neq0$, and/or $c\neq 0$, $h^{\mu}$ and $q^{\mu}$ will differ from  Eqs.~\eqref{equappenE8} and \eqref{equappenE9}. In such a situation, one will get some additional cross terms in dissipative currents. This interesting observation was pointed out in Ref.~\cite{Hu:2022azy}. \\

In our calculation, however, we do not consider such cross effects as we argue (see discussion below) that the cross effects can give rise to contradictory results (at least for zero baryon chemical potential considered herein). We start with the hydrodynamic equation,  
\begin{align}
    (\varepsilon+p)Du^{\mu}=\nabla^{\mu}p.
    \label{equappenE10}
\end{align}
One may take into account the dissipative hydrodynamic equations but that would only give rise to higher-order terms in the subsequent discussions. From  Eq.~\eqref{equappenE10} we find, 
\begin{align}
    Du^{\mu} & =\frac{\nabla^{\mu}p}{\varepsilon+p}=\frac{\partial p}{\partial T}\frac{\nabla^{\mu}T}{\varepsilon+p}+\frac{\partial p}{\partial \omega_{\alpha\beta}}\frac{\nabla^{\mu}\omega_{\alpha\beta}}{\varepsilon+p}\nonumber\\
    & =\frac{\partial p}{\partial T}\frac{\nabla^{\mu}T}{\varepsilon+p}+S^{\alpha\beta}\frac{\nabla^{\mu}\omega_{\alpha\beta}}{\varepsilon+p}
\end{align}
Note that $S^{\mu\nu}\sim \mathcal{O}(1)$, and $\omega^{\alpha\beta}\sim \mathcal{O}(\partial)$. Then the second term in the above equation is $\mathcal{O}(\partial^2)$, but the first term is $\mathcal{O}(\partial)$. Therefore, we can ignore the second term in the above equation. This implies,
\begin{align}
    Du^{\mu} & =\frac{\partial p}{\partial T}\frac{\nabla^{\mu}T}{\varepsilon+p}+\mathcal{O}(\partial^2)= s\frac{\nabla^{\mu}T}{\varepsilon+p}+\mathcal{O}(\partial^2)\nonumber\\
    & = \beta \frac{\varepsilon+p-\omega^{\alpha\beta}S_{\alpha\beta}}{\varepsilon+p}\nabla^{\mu}T+\mathcal{O}(\partial^2)\nonumber\\
    & = \beta \nabla^{\mu}T+\mathcal{O}(\partial^2)\nonumber\\
    \implies & Du^{\mu}-\beta \nabla^{\mu}T=0+\mathcal{O}(\partial^2).
    \label{equappenE11}
\end{align}
Therefore, $X^{\mu}\equiv  \beta\nabla_{\mu}T-Du_{\mu}$ can not be fixed at $\mathcal{O}(\partial)$ if we consider the on-shell conditions (hydrodynamic equations). We can use this observation in the entropy current analysis. Recall Eq.~\eqref{equappenE1}, 
\begin{align}
\partial_{\mu}\mathcal{S}^{\mu}_{\text{can} (h,q)} &=-\beta h^{\mu}\left(\beta\nabla_{\mu}T-Du_{\mu}\right)\nonumber\\
& ~~~~-\beta q^{\mu}\left(\beta \nabla_{\mu}T+Du_{\mu}-4\omega_{\mu\nu}u^{\nu}\right)\nonumber\\
& = -\beta h^{\mu}X_{\mu}-\beta q^{\mu}Y_{\mu}.
\label{equappenE12}
\end{align}

\begin{enumerate}
\item \textbf{Entropy current analysis without cross-effects:}
This is the approach that we followed in our calculation. In this case, we have, 
\begin{align}
& h^{\mu}=aX^{\mu},~ a\geq 0, \label{equappenE13} \\
& q^{\mu}=bY^{\mu}, ~b\geq 0. \label{equappenE14}
\end{align}
If we use the hydrodynamic equations, i.e., Eq.~\eqref{equappenE11}, we can argue that $X^{\mu}= 0+\mathcal{O}(\partial^2)$. If we use this observation in Eq.~\eqref{equappenE12}, then we find
\begin{align}
\partial_{\mu}\mathcal{S}^{\mu}_{\text{can} (h,q)} &=-\beta q^{\mu}Y_{\mu}+\mathcal{O}(\partial^3).
\label{}
\end{align}
The above equation again implies that $X^{\mu}$ or $h^{\mu}$ vanish at the order $\mathcal{O}(\partial)$. The terms $h^{\mu}$  or $X^{\mu}$ can only contribute to the entropy current at higher order in gradient. The same conclusion can be obtained from Eq.~\eqref{equappenE13}, namely $h^{\mu}=0+\mathcal{O}(\partial^2)$, because $X^{\mu}\sim 0+\mathcal{O}(\partial^2)$. Therefore, in this approach, the results obtained for $h^{\mu}$ using Eqs.~\eqref{equappenE12} and \eqref{equappenE13} are consistent with each other. Similarly, (from Eqs.~\eqref{equappenE12} and \eqref{equappenE14}) we can consistently find $q^{\mu}$ at $\mathcal{O}(\partial)$.  

\item \textbf{Entropy current analysis with cross-effects:} This is the approach considered in Ref.~\cite{Hu:2022azy}. We argue that this approach can lead to a logical contradiction at zero chemical potential. In this case, $h^{\mu}$ and $q^{\mu}$ are expressed as (Eqs.~\eqref{equappenE2} and \eqref{equappenE3} )
\begin{align}
    h^{\mu}& = aX^{\mu}+bY^{\mu}, \label{equappenE17}\\
    q^{\mu} & = cX^{\mu}+dY^{\mu} \label{equappenE18}. 
\end{align}
As we argued above, the positivity of the entropy production gives the following conditions,
\begin{align}
    a\beta\geq 0, \quad \beta d\geq 0, \quad 4ad\beta^2\geq (b+c)^2\beta^2.
\end{align}
Once again, let us now use the hydrodynamic equation, i.e., Eq.~\eqref{equappenE11} that gives $X^{\mu}=0+\mathcal{O}(\partial^2)$. Therefore, Eq.~\eqref{equappenE12} again leads to
\begin{align}
\partial_{\mu}\mathcal{S}^{\mu}_{\text{can} (h,q)}=-\beta q_{\mu}Y^{\mu}+\mathcal{O}(\partial^3),
\label{equappenE20}
\end{align}
that immediately implies $h^{\mu}=0+\mathcal{O}(\partial^2)$. However, we can find $q^{\mu}$ at the order $\mathcal{O}(\partial)$. Now let us look at Eq.~\eqref{equappenE17}. This equation along with Eq.~\eqref{equappenE11} implies
\begin{align}
h^{\mu}=bY^{\mu}+\mathcal{O}(\partial^2)\sim \mathcal{O}(\partial). 
\end{align}
However, this is in contradiction with Eqs.~\eqref{equappenE20} from which we concluded that 
$h^{\mu}=0+\mathcal{O}(\partial^2)$. 
The source of this contradiction is the way Eq.~\eqref{equappenE17} and $\eqref{equappenE18}$ are written. In these equations, we have expressed $h^{\mu}$ and $q^{\mu}$ as a combination of different dissipative currents. If we did not then there is no contradiction, e.g., for $b=0$, the second term in Eq.~\eqref{equappenE17} will drop, and in that case, the contradiction does not appear. This indicates that we should consider different dissipative currents individually at least for zero baryon chemical potential. 
\end{enumerate}

In order to argue that dissipative currents should be considered individually we make use of the hydrodynamic equation explicitly. Now we show that our calculation as described in Sec.~\eqref{sec:2} remains unchanged even if we use the hydrodynamic equations explicitly. 
To clearly demonstrate this, let us start with Eq.~\eqref{equ27pap1}, 
 \begin{align} \partial_{\mu}\mathcal{S}^{\mu}_{\text{can}}= &-\beta h^{\mu}\left(\beta \nabla_{\mu}T-Du_{\mu}\right)+\beta\pi^{\mu\nu}\sigma_{\mu\nu}+\beta\Pi \theta\nonumber\\ & -\beta q^{\mu}\left(\beta \nabla_{\mu}T+Du_{\mu}-4 \omega_{\mu\nu}u^{\nu}\right)\nonumber\\ & +\phi^{\mu\nu}\left(\Omega_{\mu\nu}+2\beta \Delta^{\alpha}_{~\mu}\Delta^{\beta}_{~\nu}\omega_{\alpha\beta}\right)\nonumber\\ & +2\beta\left[2k_{\alpha}\delta q^{\alpha}+\lambda_{\alpha\beta}\delta\phi^{\alpha\beta}\right]. 
\end{align}
As we have already shown, if we consider hydrodynamic equations, then $h^{\mu}\sim 0+\mathcal{O}(\partial^2)$. This observation has already been pointed out in Ref.~\cite{Hattori:2019lfp}. But in the main text for the sake of generality, we have reported the general expression of $h^{\mu}$ obtained using the entropy current analysis.  
If we use $h^{\mu}\sim 0+\mathcal{O}(\partial^2)$ in the above equation, we get
 \begin{align} \partial_{\mu}\mathcal{S}^{\mu}_{\text{can}}= &\beta\pi^{\mu\nu}\sigma_{\mu\nu}+\beta\Pi \theta\nonumber\\ & -\beta q^{\mu}\left(\beta \nabla_{\mu}T+Du_{\mu}-4 \omega_{\mu\nu}u^{\nu}\right)\nonumber\\ & +\phi^{\mu\nu}\left(\Omega_{\mu\nu}+2\beta \Delta^{\alpha}_{~\mu}\Delta^{\beta}_{~\nu}\omega_{\alpha\beta}\right)\nonumber\\ & +2\beta\left[2k_{\alpha}\delta q^{\alpha}+\lambda_{\alpha\beta}\delta\phi^{\alpha\beta}\right]+\mathcal{O}(\partial^3). 
\end{align}
Since we have already argued that cross-effects in identifying dissipative currents can be misleading at zero chemical potential, we will consider dissipative currents individually. If we follow the steps discussed in the main text, we obtain a unique constitutive relation for various dissipative currents
\begin{align} \pi^{\mu\nu}&=2\eta\sigma^{\mu\nu},\label{}\\ \Pi &= \zeta \theta,\label{}\\ q^{\mu}&=\lambda \left(\beta\nabla^{\mu}T+Du^{\mu}-4\omega^{\mu\nu}u_{\nu}\right),\label{}\\ \phi^{\mu\nu}&=\gamma\left(\Omega^{\mu\nu}+2\beta \Delta^{\mu}_{~\alpha}\Delta^{\nu}_{~\beta}\omega^{\alpha\beta}\right),\label{}
\end{align}
together with the constraint
\begin{align}
2 k_{\alpha}\delta q^{\alpha}+ \lambda_{\alpha\beta}\delta\phi^{\alpha\beta}\geq 0.
\label{}
\end{align}
Therefore, even for $h^{\mu}=0+\mathcal{O}(\partial^2)$, the dissipative currents $\pi^{\mu\nu}$, $\Pi$, $q^{\mu}$, and $\phi^{\mu\nu}$ can be uniquely obtained in terms of hydrodynamic variables, $T$, $u^{\mu}$, and $\omega^{\mu\nu}$. However, once again, the difficulty arises due to the last constraint that must be satisfied to argue that away from equilibrium the entropy should be produced. This additional condition is independent of $h^{\mu}$. Hence, even setting $h^{\mu}=0$, does not remove this additional constraint.
Therefore, we have to improve the naive canonical framework to find consistent constitutive relations for various dissipative currents.

\section{DECOMPOSITION OF $\delta T^{\mu\nu}_{\rm ph(1)}$}
\label{appenC}
%
Here we provide details on the decomposition of the first-order gradient term, the term $\delta T^{\mu\nu}_\text{ph(1)}=\partial_{\lambda}\left(u^{\nu}S^{\mu\lambda}\right)$ into different irreducible tensor structures.
This can be done by first decomposing $\delta T^{\mu\nu}_\text{ph(1)}$ into symmetric and antisymmetric parts
\begin{align}
    \delta T^{\mu\nu}_\text{ph(1)} & =\delta T^{(\mu\nu)}_\text{ph(1)}+\delta T^{[\mu\nu]}_\text{ph(1)},
    \label{equ62pap2}
\end{align}
which can be decomposed further in the following way 
\begin{align}
    & \delta T^{(\mu\nu)}_\text{ph(1)} = \delta\overline{\varepsilon} u^{\mu}u^{\nu}+\delta\overline{h}^{\mu}u^{\nu}+\delta\overline{h}^{\nu}u^{\mu}+\delta\overline{\tau}^{\mu\nu}\label{equ63pap2}\\
    & \delta T^{[\mu\nu]}_\text{ph(1)}=\delta\overline{q}^{\mu}u^{\nu}-\delta\overline{q}^{\nu}u^{\mu}+\delta\overline{\phi}^{\mu\nu}\label{equ64pap2}, 
\end{align}
along with the conditions $\delta\overline{h}^{\mu}u_{\mu}=0, ~\delta\overline{\tau}^{\mu\nu}u_{\mu}=0,~\delta\overline{q}^{\mu}u_{\mu}=0,~ \delta\overline{\phi}^{\mu\nu}u_{\mu}=0,~\delta\overline{\tau}^{\mu\nu}=\delta\overline{\tau}^{\nu\mu}, ~\delta\overline{\phi}^{\mu\nu}=-\delta\overline{\phi}^{\nu\mu}$. 

The symmetric tensor $\delta\overline{\tau}^{\mu\nu}$ can be further decomposed into a traceless part and a trace
\begin{align}
    \delta\overline{\tau}^{\mu\nu}& =\bigg( \Delta^{\mu}_{(\alpha}\Delta^{\nu}_{\beta)} -\frac{1}{3}\Delta^{\mu\nu}\Delta_{\alpha\beta}\bigg) \delta\overline{\tau}^{\alpha\beta}
    +\frac{1}{3}\Delta^{\mu\nu}\Delta_{\alpha\beta}  \delta\overline{\tau}^{\alpha\beta}\nonumber\\
    & = \delta\overline{\tau}^{\langle\mu\nu\rangle}+\delta\overline{\tau}\Delta^{\mu\nu}.
    \label{equ65pap2}
\end{align}
Note that $\delta\overline{\phi}^{\mu\nu}$ is antisymmetric, therefore it is traceless. 

The unknown quantities in Eqs.~\eqref{equ63pap2} and~\eqref{equ64pap2} can be obtained by taking the following projections
\begin{align}
    & \delta\overline{\varepsilon}=u_{\mu}u_{\nu}\partial_{\lambda}(u^{(\nu}S^{\mu)\lambda})= u_{\mu}\partial_{\lambda}S^{\mu\lambda},
    \label{equ68pap2}
\end{align}
\begin{align}
    & \delta\overline{h}^{\beta}=\Delta^{\beta}_{~\nu}u_{\mu}\partial_{\lambda}(u^{(\nu}S^{\mu)\lambda})
    = \frac{1}{2}u_{\mu}(\partial_{\lambda}u^{\beta})S^{\mu\lambda}+\frac{1}{2}\Delta^{\beta}_{~\nu}\partial_{\lambda}S^{\nu\lambda},
    \label{equ69pap2}
\end{align} 
\begin{align}    
     \delta\overline{\tau}^{\langle\mu\nu\rangle}=&\bigg( \Delta^{\mu}_{(\alpha}\Delta^{\nu}_{\beta)} -\frac{1}{3}\Delta^{\mu\nu}\Delta_{\alpha\beta}\bigg)\partial_{\lambda}(u^{(\beta}S^{\alpha)\lambda})\nonumber\\
         =&\bigg( \Delta^{\mu}_{(\alpha}\Delta^{\nu}_{\beta)} -\frac{1}{3}\Delta^{\mu\nu}\Delta_{\alpha\beta}\bigg)\partial_{\lambda}(u^{\alpha}S^{\beta\lambda}),
    \label{equ70pap2}
\end{align}
\begin{align}
     \delta\overline{\tau} = \frac{1}{3}\Delta_{\mu\nu}\partial_{\lambda}(u^{(\nu}S^{\mu)\lambda}) = \frac{1}{3}\Delta_{\mu\nu}\partial_{\lambda}(u^{\nu}S^{\mu\lambda}),
     \label{equ71pap2}
\end{align}
\begin{align}
     \delta\overline{q}^{\beta}=u_{\nu}\Delta^{\beta}_{~\mu}\partial_{\lambda}(u^{[\nu}S^{\mu]\lambda}),
     \label{equ72pap2}
\end{align} 
\begin{align}
& \delta\overline{\phi}^{\alpha\beta}= \Delta^{[\alpha}_{~\mu}\Delta^{\beta]}_{~\nu} \partial_{\lambda}(u^{[\nu}S^{\mu]\lambda}) = \Delta^{[\alpha}_{~\mu}\Delta^{\beta]}_{~\nu}\partial_{\lambda}(u^{\nu}S^{\mu\lambda}).
\label{equ73pap2}
\end{align} 
Therefore, using Eqs.~\eqref{equ62pap2}, \eqref{equ63pap2} and \eqref{equ64pap2}, the canonical energy-momentum tensor takes the form \eqref{newTcan} where $\delta$ contributions in Eqs.~\eqref{c1}-\eqref{c5} are calculable from Eqs.~\eqref{equ68pap2}-\eqref{equ73pap2}.  



 \bibliographystyle{elsarticle-num} 
 \bibliography{cas-refs}





\end{document}